\documentclass[authoryear,round,12pt]{article}
\usepackage{color}
\usepackage{pslatex}
\usepackage{graphicx}
\usepackage{setspace}
\usepackage{natbib}
\usepackage{amsmath}
\usepackage{subfigure}
\newcommand{\dg}{$^\circ$}

\newcommand{\beq}{\begin{equation}}
\newcommand{\eeq}{\end{equation}}
\newcommand{\myabstract}{This is part of a series of weekly influenza
  forecasts made during the 2012-2013 influenza season.  Here we
  present results of forecasts initiated following assimilation of
  observations for Week 1 (i.e. the forecast begins January 6,
  2013) for municipalities in the United States.  These forecasts were
  performed on January 11, 2013.  Results from forecasts initiated the
  six previous weeks (Weeks 47-52) are also presented.

  The accuracy of these predictions will not be known for certain
  until the conclusion of the current influenza season; however, at
  the moment a number of the forecasted peaks appear to be inaccurate.
  This inaccuracy may be due to the virulence of influenza this
  season, which appears to be sending more influenza-infected persons
  to seek medical attention and inflates ILI levels (and possibly the
  proportion testing influenza positive) relative to years with milder
  flu strains.  New forecasts that adjust, or scale, for this
  difference and match the two focus cities that appear to have
  already peaked are identified.  These new forecasts will be used, in
  addition to the previously scaled forms, to make influenza
  predictions for the remainder of the season. }
\begin{document}
%
%%%%%%%%%%%%%%%%%%%%%%%%%%%%%%%%%%%%%%%%%%%%%%%%%%%%%%%%%%%%%%%%%%%%%
% TITLE
%
% Enter your TITLE here
%%%%%%%%%%%%%%%%%%%%%%%%%%%%%%%%%%%%%%%%%%%%%%%%%%%%%%%%%%%%%%%%%%%%%
\title{\textbf{\large{Week 1 Influenza Forecast for the 2012-2013
      U.S. Season}}}
%
% Author names, with corresponding author information. 
% [Update and move the \thanks{...} block as appropriate.]
%
\author{\textsc{Jeffrey Shaman}
                                \thanks{\textit{Corresponding author address:} 
                                Jeffrey Shaman, Department of
                                Environmental Health Sciences, Mailman
                                School of Public Health, Columbia
                                University, 722 West 168th Street,
                                Rosenfield Building, Room 1104C, New
                                York, NY 10032. 
                                \newline{E-mail:
                                  jls106@columbia.edu}}\quad\textsc{}\\
\centerline{\textit{\footnotesize{Department of Environmental Health Sciences,
    Mailman School of Public Health, Columbia University, New York, New York}}}
\and
\centerline{\textsc{Alicia Karspeck}} \\% Add additional authors, different insitution
\centerline{\textit{\footnotesize{Climate and Global Dynamics
      Division, National Center for Atmospheric Research, Boulder, Colorado}}}
\and 
\centerline{\textsc{Marc Lipsitch}} \\% Add additional authors, different insitution
\centerline{\textit{\footnotesize{Center for Communicable Disease
      Dynamics, Harvard School of Public Health, Harvard University,
      Boston, Massachussetts}}}
}

\maketitle

{
%\amstitle
\begin{abstract}
\myabstract
\end{abstract}
}
%%%%%%%%%%%%%%%%%%%%%%%%%%%%%%%%%%%%%%%%%%%%%%%%%%%%%%%%%%%%%%%%%%%%%
% MAIN BODY OF PAPER
%%%%%%%%%%%%%%%%%%%%%%%%%%%%%%%%%%%%%%%%%%%%%%%%%%%%%%%%%%%%%%%%%%%%%

\section{Observation Bias and Scaling}
\label{sec:obs}

It is becoming clear that many of the forecasts archived in previous
weeks predicted outbreak peaks that are too early.  In part, the model
grew overly confident in its own consensus or spread, but also the
forecasts had difficulty handling the high levels of influenza
incidence this year.  Understanding why the forecasts are predicting
peaks earlier than observed should help improve predictions going
forward, both this year and in the future.

There are 3 components to the system that enable training of the model
prior to forecast: the model itself, the observations and the data
assimilation method.  Here we will focus on the observations we have
been using, which we term ILI+.  There are at least 3 issues with the
ILI+ data, which we've thus far identified, that may be corrupting
model training and potentially degrading forecast accuracy.  These
are:

\begin{enumerate}
\item Interseasonal differences in influenza virulence, defined here
  as the probability an influenza-infected person seeks medical
  attention
\item Intraseasonal biases that can arise in Google Flu Trends
  estimates of ILI
  \item Under-reporting and biases of CDC regional influenza positive
    rates when first posted.
\end{enumerate}

Each of these issues are discussed in turn herewith.

\subsection{Influenza Virulence and ILI}
\label{subsec:virul}

For these forecasts we using a metric we term ILI+, which is weekly
municipal level Google Flu Trends (GFT) ILI multiplied by the
proportion of ILI patients testing influenza positive at the CDC
regional census division level.  The GFT ILI estimate
\citep{Ginsberg-Mohebbi-Patel-et-al-2009:detecting} is calibrated to
match CDC ILI estimates, which is the number of people with ILI per
100,000 patient visits in the U.S.

Multiplying the GFT ILI times the census division proportion influenza
positive gives an estimate of number of flu cases per 100,000 patient
visits.  If the strain(s) in a given flu season are not virulent and
symptoms are weak, then there will many more cases in the population
than seen in sentinel clinics and reported in either ILI or ILI+.
That is, many people will get the flu, experience mild or no symptoms,
and never show up in a sentinel clinic or hospital, such that for
every reported ILI ``person'' there are many with mild symptoms who
never have a patient visit.  That suggests that for mild flu, the
actual number of flu infections per 100,000 \textit{persons} (not
patients) is higher than the ILI or ILI+ estimate.

On the other hand, if the strain is more virulent, as it appears to be
this year, then the ILI and ILI+ metrics should be \textit{more}
representative of rates of flu incidence in the general population.
How much more representative is very difficult to gauge, because there
is actually an additional factor needed to convert weekly ILI+ per
100,000 patient visits to weekly influenza incidence per 100,000
people--the overall rate at which a population seeks medical
attention.  That is, two probabilities need to be constrained:

\begin{enumerate}
\item the probability that a person with influenza seeks medical
  attention, i.e. the virulence factor
\item the probability that anyone seeks medical attention for any
  reason, i.e. the number of patient visits per 100,000 people
\end{enumerate}

The first factor converts ILI+ to total influenza.  The second factor
converts the denominator from patient visits to a total population
measure.  The combined effects of these two factors, sometimes
referred to as the ``syndromic multiplier'', has had only limited
assessment \citep[e.g.][]{metzger-hajat-crawford-et-al-2004:many}.

However, given the virulence of the flu this year it seems sensible
that the ILI and ILI+ metrics are more representative of the general
population than in years when the symptoms are mild and fewer persons
with influenza are inclined to see a doctor.

Indeed, for most of the last decade influenza has not been
particularly virulent.  Prior to running these real-time forecast, we
ran retrospective forecasts with an accounting for the
``representativeness'' of ILI for the general population.  That is,
the ILI+ raw number was multiplied by a factor to account for all the
mild and asymptomatic infections not detected by current surveillance,
as well as the number of patient visits per 100,000 persons per week.
Scaling factors of 4-10 seemed to produce the best predictions in
retrospect for the 2003-2004 through 2011-2012 seasons.  For some
seasons factors of 2.5 or 15 or 20 even worked well.

Given the retrospective forecast performance, we went with a factor of
5 scaling of the ILI+ for the forecasts we've been archiving in real
time \citep{Shaman-Karspeck-Lipsitch-2012:week49,
  Shaman-Karspeck-Lipsitch-2012:week50,
  Shaman-Karspeck-Lipsitch-2012:week51,
  Shaman-Karspeck-Lipsitch-2013:week52}--though we have been running
some alternately scaled forecasts in real time as well.  That is, a
factor of 5 scaling of the ILI+ more or less produced the best
retrospective predictions on average for the 2003-2004 through
2011-2012 seasons.

However, given that these forecasts using ILI+ multiplied by a factor
of 5 appear to have produced poor predictions (see below), and that
this year's flu is indeed virulent, it seems appropriate to consider
whether the forecasts should be run with a lower factor.  This
alternative is explored in Section \ref{subsec:actual1}.

\subsection{Intraseasonal Biases in Google Flu Trends}

GFT ILI estimates are calibrated to the CDC ILI estimates derived from
the sentinel provider surveillance system, ILInet.  The GFT ILI
estimates are derived from internet search activity, specifically
search queries, such as ``influenza complication''.  Changes in online
health-seeking behavior during an outbreak can affect the quality of
GFT ILI estimates \citep{cook-conrad-fowlkes-et-al-2011:assessing}.

One concern for the current season is that the widespread media
attention given to the influenza outbreak--e.g. influenza has been the
lead story on national nightly news programs, the declaration of a
public health emergency in Boston in response to the outbreak,
etc.--since early December, may be affecting online search behavior.
We might expect that with the increased attention and awareness of
influenza, health-seeking search activity would increase and GFT ILI
estimates would become elevated relative to CDC ILInet estimates.

\begin{figure}[tbh]
   \noindent\includegraphics[width=20pc,angle=0]{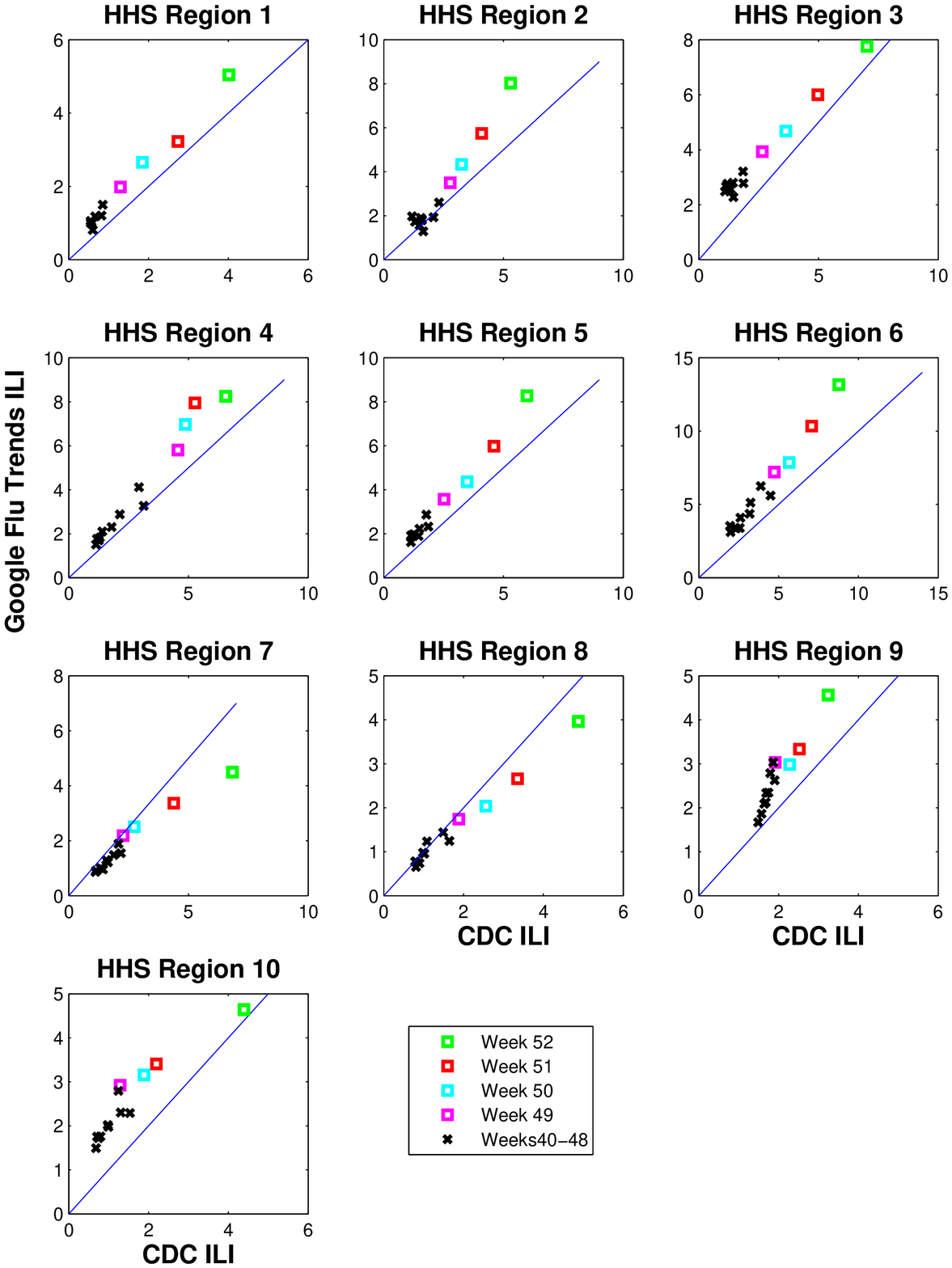}
   \noindent\includegraphics[width=20pc,angle=0]{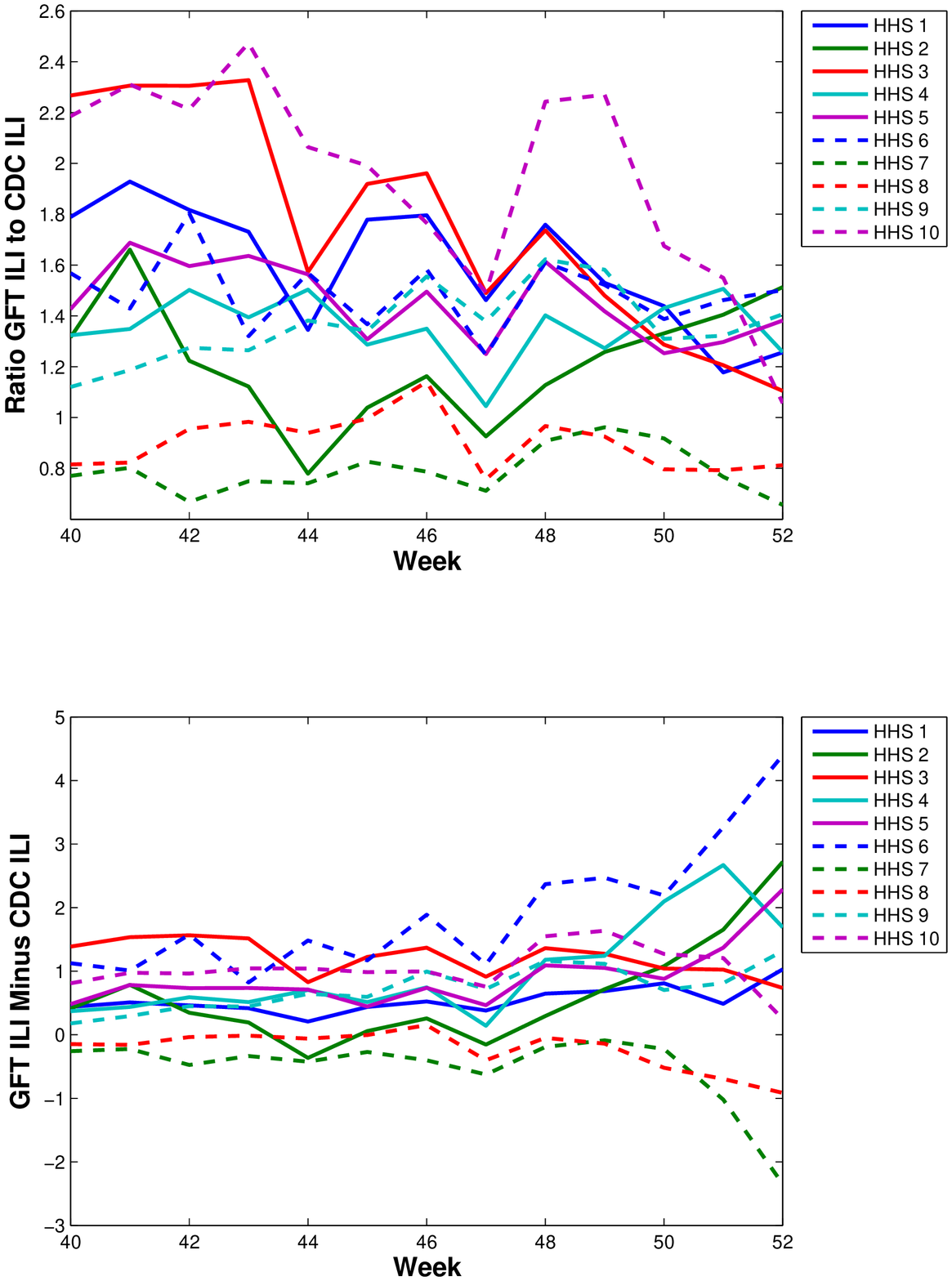}\\
 \caption{Left) Scatter plots of weekly GFT ILI plotted as a function of the
   corresponding CDC ILInet estimate. Units are percent ILI among
   patient visits.   Weeks 40-52 of the 2012-2013
   season are shown for each HHS region.  The blue line represents
   equivalent estimates.  Right) Plots of the weekly ratio of GFT ILI to CDC
   ILI through time for weeks 40-52 of the 2012-2013 season.}
 \label{fig:gft_bias} 
\end{figure}

The evidence for bias at the regional level is given in Figure
\ref{fig:gft_bias}. \textit{This is a correction from the initial
  posting of this article, which used the raw numerator ILI values
  from the CDC ILInet estimates, erroneously assuming those values
  were already adjusted to reflect ILI per 100,000 patient visits (JS
  error).}

In the best case, plots of GFT ILI estimates as a function of CDC
ILInet estimates should fall on the 45\dg diagonal (blue line), which
indicates an equivalence of the two measures.  This correspondence is
evident early in the season for HHS regions 2, 4, 5, 7 and 8.  A few
of the other HHS regions (3, 6 and 10) appear to lie near parallel to
this line early in the season, but a bit above it, indicating the GFT
ILI estimates are biased high.  The early season portion for Region 9
has a scatter plot that is linear, but not parallel to the blue line;
that is, the bias increases with more ILI.

Of greater concern would be any non-linear component to the
scatterplots--i.e. an indication that the bias is neither additive or
multiplicative.  There is some evidence of this for HHS regions 2, 4,
5, and 7, which indicate an increase in bias during December,
i.e. within season/through time, when public attention on influenza
intensified.  For all these, but Region 7, the change increases the
GFT ILI overestimate of the CDC ILI.  Another way to view the change
in bias over the season is with plots the ratio of GFT ILI to CDC ILI
and the difference of GFT ILI minus CDC ILI versus time (Figure
\ref{fig:gft_bias} right).  These show that while the absolute
distance between the two metrics has increased in recent weeks (up to
a factor of 4 through Week 52 for Region 7), the ratio has not changed
substantially.

For most of the HHS regions, the GFT ILI is a bit high; however, the
changes to biases within season appear not nearly as problematic as
previously (erroneously) posted.  The analysis still remains
preliminary.  It is possible that the CDC ILInet numbers will continue
to be updated and that biases will change.

\subsection{Shifting Influenza Positive Rates}

The third problem with the ILI+ data stems from updates to CDC posted
weekly influenza positive rates subsequent to their initial posting.
The proportion influenza positive still shows a propensity to increase
from first posting to second.  For weeks 47-52, we compared reports
from the 9 census divisions (54 total week-census division
combinations), each posted first 6 days following the given week, then
updated 13 days following that week.  Of these 54 week-census division
combinations, the proportion influenza positive increased 46 times
from the first to second posting, and decreased only 8 times.

The null hypothesis is that for a representative first-posted
proportion influenza positive, there should be an equal chance of the
updated rate being higher or lower.  Bootstrapping this, there is less
than a 0.00005\% probability ($p<0.0000005$) that 46 of 54 instances
would increase (or decrease) by chance alone.  It appears that there
is some systematic tendency for later receipt of influenza positive
assays, as if these positive assays take longer to report.  Perhaps
there is an additional confirmation test performed that delays them.

Table \ref{table:ta} shows the percent testing influenza positive for
Weeks 51 and 52 (18 of the total 54 week-census division
combinations), as posted initially 6 days after the end of the week,
and then again (updated) 13 days after the end of the week.  The
influenza positive rates of only 2 of the 18 initial postings decrease
with the subsequent updating.

\begin{table}[t]
  \caption{Percentage testing influenza positive at the Census
    Division level for Weeks 51 and 52, as first posted and
    subsequently updated one week later.  The additive change from the first
    posting to the second is also presented. }\label{table:ta}
 \begin{center}
\resizebox{16cm}{!} {
 \begin{tabular}{|c|c|c|c|c|c|c|c|}
   \hline
  &\multicolumn{3}{|c|}{Week 51}&\multicolumn{3}{|c|}{Week52}\\
\hline
  Census Division & Dec. 28, 2012 & Jan. 4, 2013 &  Change&Jan. 4 2013 & Jan. 11, 2013 & Change\\
& Posting & Posting && Posting & Posting &\\
   \hline
Northeast&27.87\%&49.88\%&\textcolor{red}{+22.01\%}&37.75\%&45.76\%&\textcolor{red}{+8.01\%}\\
Mid-Atlantic & 20.04\%& 40.11\% &
\textcolor{red}{+20.07\%}&31.65\%&31.60\%&\textcolor{blue}{-0.05\%}\\
East North Central
&61.97\%&56.73\%&\textcolor{blue}{-5.24\%}&50.15\%&53.26\%&\textcolor{red}{+3.11\%}\\
West North Central & 25.95\%& 42.28\% & \textcolor{red}{+16.33\%} &
30.53\%&45.47\% &\textcolor{red}{+14.94\%}\\
South Atlantic & 29.94\%& 34.01\%& \textcolor{red}{+4.07\%} & 27.11\%
&31.33\% &\textcolor{red}{+4.22\%}\\
East South Central & 26.61\% &38.56\%& \textcolor{red}{+11.95\%}&
30.28\% & 42.06\% &\textcolor{red}{+11.78\%}\\
West South Central & 13.23\% &26.49\%& \textcolor{red}{+13.26\%}&
20.23\% & 30.68\% &\textcolor{red}{+10.45\%}\\
Mountain & 36.83\% &39.45\%& \textcolor{red}{+2.62\%}&
39.16\% & 39.82\% &\textcolor{red}{+0.66\%}\\
Pacific & 15.54\% &16.68\%& \textcolor{red}{+1.14\%}&
20.42\% & 22.90\% &\textcolor{red}{+2.48\%}\\
\hline
 \end{tabular}
}
 \end{center}
\end{table}

\begin{figure}[tbh]
\noindent\includegraphics[width=20pc,angle=0]{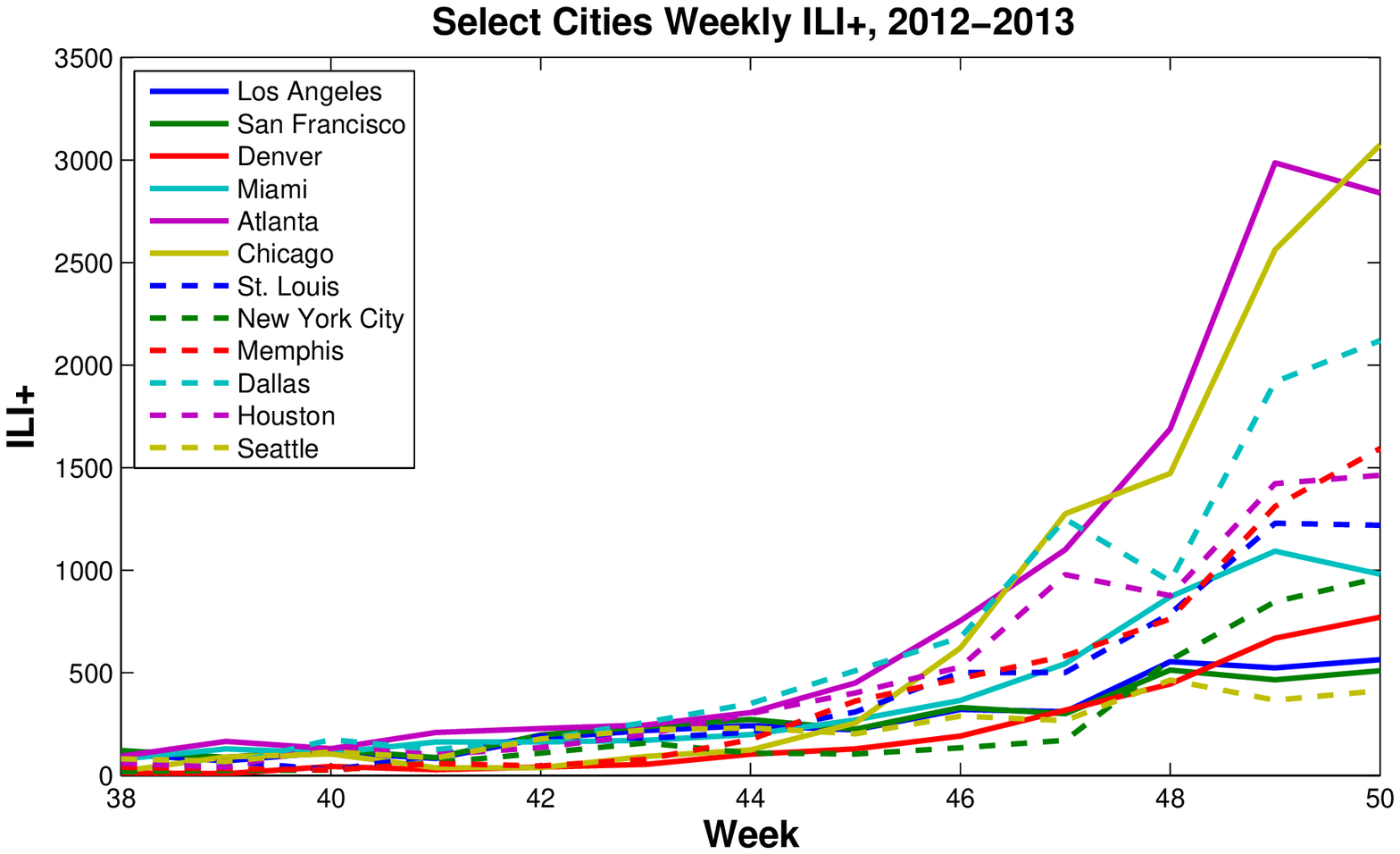}
\noindent\includegraphics[width=20pc,angle=0]{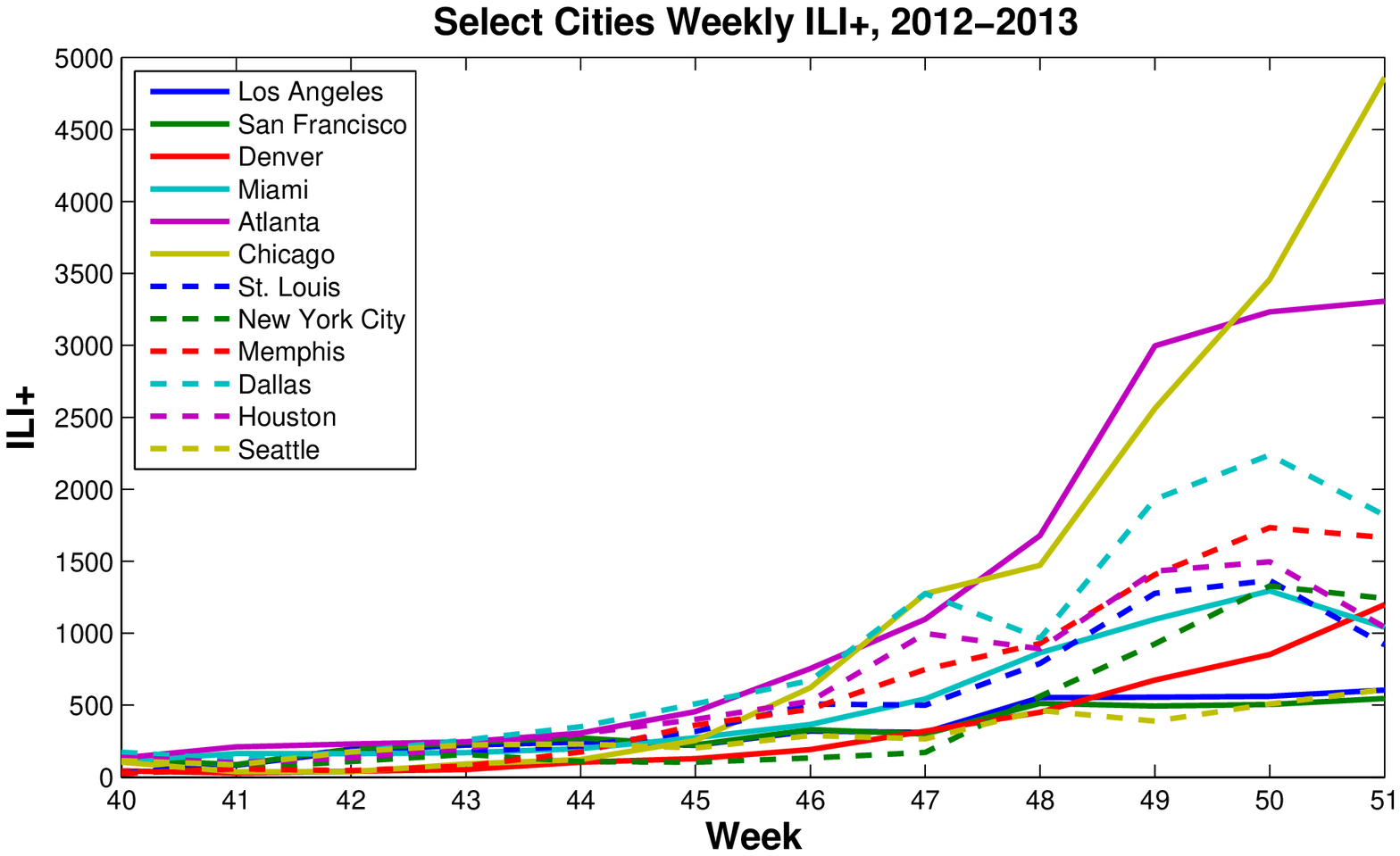}\\
\noindent\includegraphics[width=20pc,angle=0]{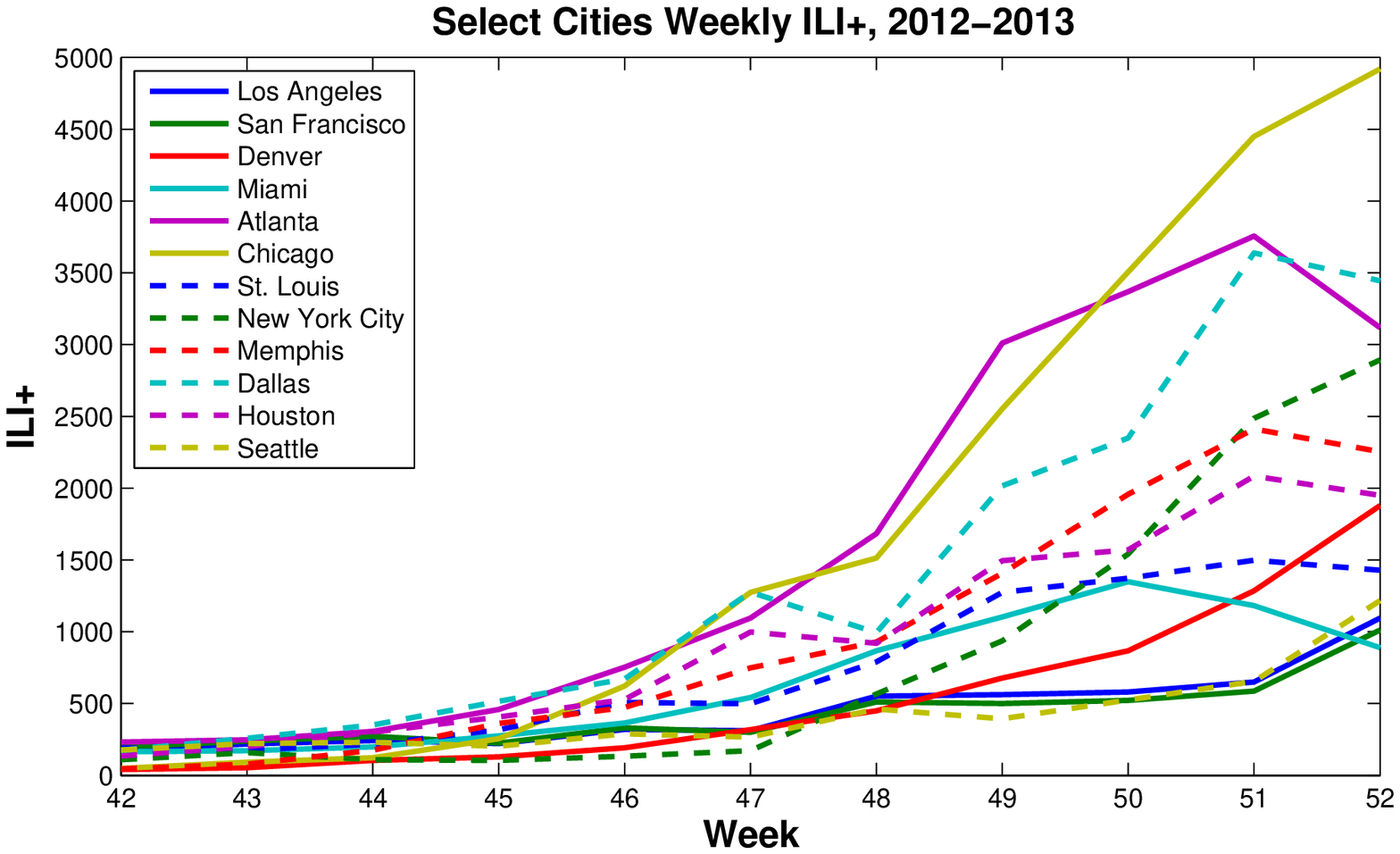}
\noindent\includegraphics[width=20pc,angle=0]{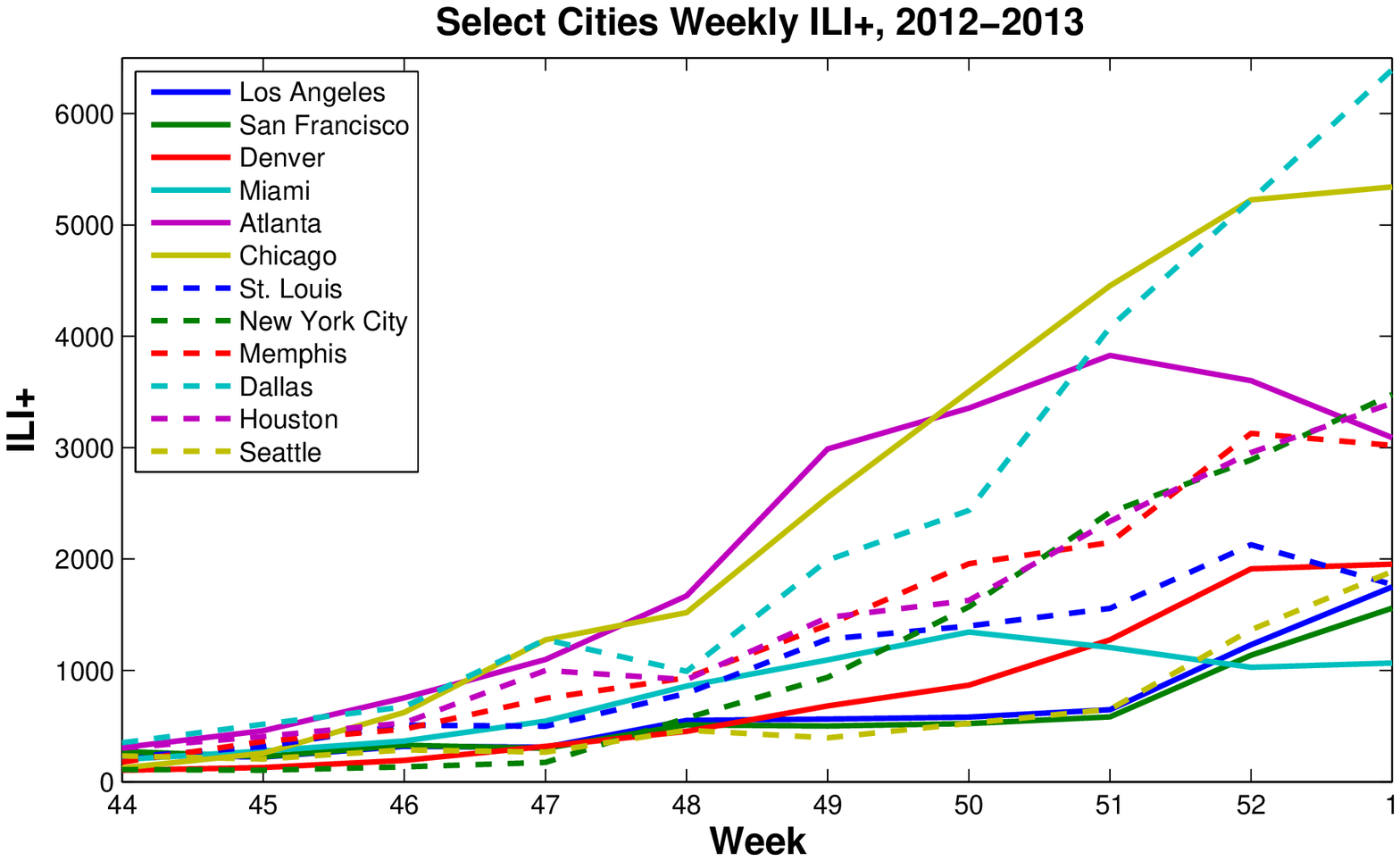}
\caption{Time series of: Top Left) Week 50 estimates of Weeks 38-50
  ILI+; Top Right) Week 51 estimates of Weeks 40-51 ILI+; and Bottom
  Left) Week 52 estimates of Weeks 42-52 ILI+; Bottom Right) Week 1
  estimates of Weeks 44-1 ILI+ for the 2012-2013 season.  ILI+ is
  Google Flu Trends weekly municipal ILI estimates times CDC census
  division seropositive rates.}
\label{fig:selcitytimeseries}
\end{figure}

The shifts can also be seen in the Figure \ref{fig:selcitytimeseries}.
Only Atlanta and Miami have held their peaks in the last few weeks.
This suggests (see below) that we can use the accuracy of the Miami
and Atlanta predictions to choose the factor of scaling for the ILI+
observations to adjust for virulence.

\section{2012-2013 Forecast}
\label{sec:actualfore}

Just to repeat: at the moment it appears some of the forecasts are
inaccurate.  (We'll be able to quantify how much so, once the season
is concluded.)  Again, one possible explanation for the inaccuracy is
due to the virulence of influenza this year, in particular the H3N2
strain.  In performing our forecasts, we used a scalling of ILI+ to
influenza incidence per 100,000 people that historically produced the
best forecasts in previous years (2003-2004 through 2011-2012);
however, as detailed above (Section \ref{subsec:virul}), this scaling
may change if the flu is more virulent and a higher percentage of the
population seeks medical attention.

\subsection{Week 1 Forecast}
\label{subsec:actual1}

We first show the forecast form we've been presenting the last 6
weeks.  Table \ref{table:t1} presents the forecasts initiated after
assimilation of observations through Week 1.  The first forecast day
is January 6, 2013.  These forecasts use the AH-forced SIRS model and
were performed on January 11, 2013 following training with ILI+
through Week 1 (using CDC census division influenza positive rates as
published online on January 11, 2013).  The ILI+ is multiplied by 5
and this scaled metric is assumed to give weekly infection incidence
per 100,000 \textit{persons}.

\begin{table}[t]
  \caption{Summary of weekly model predictions at 12 select cities.  Weeks
    are labeled consecutively (Week 1 of 2013 is Week 53, etc.).
    Predictions were initiated at the end of Weeks 47-1 and used a
    factor of 5 multiplication of the ILI+ data.
    The range of prediction confidences, derived from municipal,
    regional and national calibrations, are given in parentheses.}\label{table:t1}
 \begin{center}
\resizebox{16cm}{!} {
 \begin{tabular}{ccccrrcrc}
   \hline\hline
   City & Week 1 & Week 52 &Week 51 & Week 50 & Week 49 & Week 48 & Week 47\\
   & Prediction &Prediction &Prediction & Prediction & Prediction & Prediction & Prediction \\
   \hline
   Los Angeles &53 (80-99\%)&53 (35-95\%)&51-52 (35-60\%)&52 (50-95\%) & 51-52 (35-90\%) & 51-52 (20-55\%) & 51 (15-30\%) \\
   San Francisco &53 (80-99\%)&53 (35-60\%)&52 (25-45\%)& 52 (35-85\%) & 51-52 (25-40\%) & 51 (30-85\%) &
   50-51 (25-60\%)  \\
   Denver &53 (90-99\%)&52 (60-99\%)&52 (50-99\%)& 52 (20-60\%) & 52 (20-55\%) & 51-52 (0-55\%) & 51 (10-30\%) \\
   Miami &51 (65-99\%)&51 (65-99\%) &51-52 (30-99\%)& 51 (40-60\%) & 51 (40-99\%) & 50-51 (40-55\%) & 50-51 (0-45\%) \\
   Atlanta &49-50 (5-95\%)&50 (80-99\%)&49-50 (80-99\%)& 49 (80-99\%) & 49 (90-99\%) & 49 (80-95\%) & 49 (80-95\%) \\
   Chicago &50 (90-99\%)&50 (55-95\%)&50 (55-95\%)&49-50 (55-95\%) & 49 (55-95\%) & 49 (35-80\%) & 49 (35-80\%) \\
   St. Louis &50-51 (65-99\%)&51 (80-99\%) &50 (80-99\%)& 51 (85-99\%)  & 50-51 (80-99\%) & 50 (85-99\%) & 51 (30-90\%) \\
   New York City&52 (75-99\%)&52 (85-99\%) &51-52 (20-99\%) &52 (25-99\%) & 51 (25-99\%) & 52-53 (25-60\%) & 53-54
   (25-55\%)\\
   Memphis &50-51 (70-95\%)&51 (80-99\%)&50 (70-99\%)&51 (20-80\%) & 50 (20-80\%) & 50 (15-80\%) & 49-50 (15-55\%) \\
   Dallas &50 (65-90\%)&50 (40-70\%)&50 (65-95\%) &50 (65-90\%) & 49-50 (65-85\%) & 49 (50-75\%) & 49 (40-85\%) \\
   Houston &51 (70-95\%) & 50-51 (70-95\%) &50 (70-95\%)&50 (75-90\%) & 50 (50-60\%) & 50 (50-60\%) & 49 (50-85\%) \\
   Seattle &53-54 (80-99\%)&53 (50-90\%) &51-52 (20-60\%)&52-53 (0-55\%) & 52-53 (5-55\%) & 51-52 (5-55\%) & 51 (5-35\%) \\
   \hline
 \end{tabular}
}
 \end{center}
\end{table}

Many of the forecasts made in this fashion are off by more than 1 week
at this point (e.g. Dallas, Chicago, Houston, and partially, Memphis
and St. Louis--compare with Figure \ref{fig:selcitytimeseries}, Bottom
Right).

If the ratio of people with influenza per 100,000 persons to rates of
ILI+ per 100,000 patient visits does indeed go down as the virus
virulence goes up, then the factor of 5 that we were using to scale
the ILI+ numbers, which worked well for many of the previous
symptomatically milder outbreaks over the last 10 years, may be too
high.  That is,

\begin{equation}
\frac{\text{ILI+}/100,000 \text{ patient visits}}{\text{people with
    influenza}/100,000 \text{ persons}} \propto \text{ strain virulence}
\end{equation}

If this is so, then we need to increase the ILI+ numbers \textit{by
  less}.  Again, this is because a greater portion of those who are
infected are winding up in sentinel clinics.

We repeated the same Week 1 forecasts but with increases to the ILI+
of factors of 1, 2, 2.5, and 3 (Table \ref{table:t2}).  Note, that the
certainties given in parenthesis are now highly speculative, as there
is, at present, no appropriate historically derived confidences from
which to calibrate forecast accuracy for these alternate forms.

\begin{table}[t]
  \caption{January 11, 2013 forecasts using the AH-forced SIRS model,
    ILI+ observations and EAKF assimilation during training.  Training
    is through Week 1 (ending January 5, 2013).  ILI+ observations are
    scaled by factors of 1, 2, 3 and 5.  Certainties, based on
    historical municipal, regional and national predictions \textit{using a
    factor of 5}, are given in parentheis.  The appropriateness of using
  certainties one observational scaling to assign probabilities to
  another is questionable; however, as previous years have typically
  not been so virulent, using the correct scaling would likely be
 off as well.}\label{table:t2}
 \begin{center}
\resizebox{16cm}{!} {
 \begin{tabular}{ccccrrcrc}
   \hline\hline
   City & $1\times$ILI+ &  $2\times$ILI+ & $2.5\times$ILI+ & $3\times$ILI+ &  $5\times$ILI+ \\
  \hline
   Los Angeles&5 (10-35\%) &3-4 (30-60\%)&3 (30-60\%)&2 (30-95\%)&1 (80-99\%)\\
   San Francisco & 5-6 (20-60\%)&4 (35-60\%)&3-4 (35-56\%)&2-3 (35-60\%)&1 (80-99\%)\\
  Denver&4-5 (15-57\%) &2-3 (15-60\%)&2 (15-60\%)&1-2 (45-99\%)&1 (90-99\%)\\
   Miami &5-6 (5-35\%)&51, 3 (40-80\%)&51 (50-99\%)&51 (65-75\%)&51 (65-99\%)\\
   Atlanta &52 (1-80\%)&51 (80-90\%)&51 (80-90\%)&50-51 (5-99\%)&49-50 (5-95\%)\\
   Chicago&52 (55-80\%) &51 (55-94\%)&51 (55-95\%)&51 (55-95\%)&50 (90-99\%)\\
   St. Louis&4-5 (30-90\%) &52, 2 (35-99\%)&52 (55-99\%)&52 (65-99\%)&50-51 (65-99\%)\\
   New York City&3 (20-55\%)&1-2 (20-99\%)&1 (75-99\%)&52-1 (75-99\%)&52 (75-99\%)\\
  Memphis &3 (15-55\%)&52-1 (35-99\%)&52 (70-99\%)&52 (70-99\%)&50-51 (70-95\%)\\
   Dallas &2 (65-90\%)&52 (80-95\%)&52 (80-95\%)&52 (80-95\%)&50 (65-90\%)\\
   Houston &3 (50-85\%)&1 (70-95\%) &1 (70-99\%)& 52 (85-99\%)&51 (70-95\%)\\
   Seattle&5 (5-35\%) &3 (50-60\%)&3 (50-60\%)&2 (50-85\%)&1-2 (80-99\%)\\
   \hline
 \end{tabular}
}
 \end{center}
\end{table}

As the scaling factor is reduced, the ILI+ numbers are reduced
(similarly throughout the record), and forecasted peak week singularly
extends farther into the future.  Whether this change is appropriate
will only be evident once the season is over.  Further, we don't have
a certain means for determining which of the scaling factors is best;
however, there are 2 cities, Atlanta and Miami, that appear to be past
their ILI+ peak week (Figure \ref{fig:selcitytimeseries}, bottom
right).  That is, both cities have declined for 2 or more weeks since
their ILI+ peak during Week 50 (Miami) and 51 (Atlanta).  Therefore,
we might argue that the scaling that is most sensible to include in
future forecasts is the one that best matches those 2 peaks.  This
appears to be the 2.5 scaling, which predicts Week 51 for both, which
is spot on for Atlanta, and within the $\pm 1$ week prediction error
for Miami.  

At lower scaling factors a second larger peak is predicted for Miami
in the future.  Obviously, this could happen--we don't know.  Choosing
the scaling is guess work at this point.  Our ultimate aim will be to
develop a sensible set of rules for determing the scaling for a given
outbreak (for instance, based on strain virulence).

Table \ref{table:t3} gives the Week 47-Week 1 predictions using the
2.5 scaling factor.  Again, the confidences are based on the
retrospective 2003-2004 through 2011-2012 forecasts made with a
scaling factor of 5, and should viewed as preliminary.  They do however
provide a loose indication of the spread of the ensemble.  (Note also
that given that most, if not all, of the outbreaks between 2003-2004
and 2011-2012 were less virulent than this season, it is not clear if
retrospective calibration with a 2.5 scaling factor would give
sensible forecast certainties.)

\begin{table}[t]
  \caption{Summary of weekly model predictions at 12 select cities.  
    Predictions were initiated at the end of Weeks 47-1 and using a
    factor of \textbf{2.5} multiplication of the ILI+ data.
    The range of prediction confidences, derived from municipal,
    regional and national calibrations \textit{from retrospective runs
      using a scaling factor of 5 times ILI+}, are given in parentheses.}\label{table:t3}
 \begin{center}
\resizebox{17cm}{!} {
 \begin{tabular}{ccccrrcrc}
   \hline\hline
   City & Week 1 & Week 52 &Week 51 & Week 50 & Week 49 & Week 48 &
   Week 47\\
   & Prediction &Prediction &Prediction & Prediction & Prediction & Prediction & Prediction \\
   \hline
  Los Angeles&3 (30-60\%)&2-3 (30-60\%)&2-3 (10-45\%)& 2 (10-35\%) &
  1-2 (10-35\%) & 52-1 (10-35\%) & 52-53 (15-40\%)\\
   San Francisco &3-4 (35-56\%)&3 (35-60\%)&2-3 (20-60\%) & 2
   (20-56\%) & 1-2 (20-60\%) & 52-1 (20-55\%) &52 (15-55\%)\\
  Denver&2 (15-60\%)&1-2 (40-60\%)&1-2 (15-60\%) & 1-2 (0-60\%) &1
  (0-35\%) &52-1 (0-30\%)& 51-52 (5-40\%)\\
   Miami &51 (50-99\%)&51, 2 (35-99\%)&1 (30-60\%) &1 (30-60\%) &52
   (30-60\%) &52 (0-35\%)& 52-53 (5-30\%)\\
   Atlanta &51 (80-90\%)&51 (80-90\%)&50-51 (80-90\%)& 51 (20-99\%)&
   50 (80-99\%) & 50-51 (20-60\%) &50-51 (20-60\%)\\
   Chicago&51 (55-95\%)&51 (55-95\%)&51 (55-95\%) & 51 (40-85\%) & 50
   (40-85\%) & 50 (35-85\%) & 50-51 (30-55\%)\\
   St. Louis&52 (55-99\%)&51, 1 (35-99\%)&52-1 (35-65\%) & 52
   (35-90\%) &52 (35-90\%) & 52 (30-90\%) &52 (25-85\%)\\
   New York City&1 (75-99\%)&1 (20-99\%)&1 (20-53\%) &1 (20-55\%)&
   52-1 (25-85\%) & 52-2 (25-60\%) & 52-2 (10-50\%)\\
  Memphis &52 (70-99\%)&52 (70-90\%)&52 (15-55\%) &52 (15-85\%) & 52
  (15-85\%) & 51 (15-85\%) & 51 (0-45\%)\\
   Dallas &52 (80-95\%)&52 (80-95\%)&52 (35-65\%) &51 (50-90\%) & 51
   (50-90\%) &51 (35-75\%) & 50 950-75\%)\\
   Houston  &1 (70-99\%)& 52-1 (50-95\%)&50, 52, 1 (10-75\%) &52
   (65-85\%) & 52 (55-85\%) & 52 (10-80\%) & 50-51 (30-85\%)\\
   Seattle&3 (50-60\%)&3 (0-60\%)&3 (5-35\%) & 2-3 (5-35\%) &2-3
   (10-30\%) &52-1 (5-30\%) & 52-1 (5-30\%)\\
   \hline
 \end{tabular}
}
 \end{center}
\end{table}
 
The forecasts with the 2.5 scaling predict outbreak peaks later in the
year than the forecasts with a 5 scaling factor (compare to Table
\ref{table:t1}).  For instance, for the last 3 weekly forecasts (Weeks
51-1), Dallas is predicted to have a Week 52 rather than Week 50 peak.
Seattle has been predicted to peak during Week 3 for the last 5
forecasts with the 2.5 scaling--albeit with low certainty, but this is
to be expected given the long lead of this prediction (up to 6 weeks
lead).

Some city forecasts shift from week to week with the new scaling. With
each weekly prediction Los Angeles and San Francisco continually shift
the forecast outbreak peak about a half week later. The Chicago
prediction still fails.  It will be very interesting to evaluate these
predictions in terms of their accuracy--to tally correct predictions
based on lead time, determine if and when relative to the actual peak
a given city's forecast becomes accurate and if it remains accurate
with subsequent forecasts, and whether the forecast certainties
calibrate to overall forecast accuracy.

% Create a bibliography directory and place your .bib file there.
%\ifthenelse{\boolean{dc}}
{}
{\clearpage}
\bibliographystyle{apalike} 
\bibliography{week48bib}

\begin{thebibliography}{}

\bibitem[Cook et~al., 2011]{cook-conrad-fowlkes-et-al-2011:assessing}
Cook, S., Conrad, C., Fowlkes, A., and Mohebbi, M. (2011).
\newblock Assessing {G}oogle flu trends performance in the {United S}tates
  during the 2009 influenza virus {A (H1N1)} pandemic.
\newblock {\em PLoS ONE}, 6(8):e23610.

\bibitem[Ginsberg et~al., 2009]{Ginsberg-Mohebbi-Patel-et-al-2009:detecting}
Ginsberg, J., Mohebbi, M., Patel, R., Brammer, L., Smolinski, M., and
  Brilliant, L. (2009).
\newblock Detecting influenza epidemics using search engine query data.
\newblock {\em Nature}, 457(7232):1012--1014.

\bibitem[Metzger et~al., 2004]{metzger-hajat-crawford-et-al-2004:many}
Metzger, K., Hajat, A., Crawford, M., and Mostashari, F. (2004).
\newblock How many illnesses does one emergency department visit represent?
  {U}sing a population-based telephone survey to estimate the syndromic
  multiplier.
\newblock {\em MMWR Morb Mortal Wkly Rep}, 53(106):11.

\bibitem[Shaman et~al., 2012a]{Shaman-Karspeck-Lipsitch-2012:week49}
Shaman, J., Karspeck, A., and Lipsitch, M. (2012a).
\newblock Week 49 influenza forecast for the 2012-2013 {U.S.} season.
\newblock {\em ArXiv}, 1212.4678 [q-bio.PE].

\bibitem[Shaman et~al., 2012b]{Shaman-Karspeck-Lipsitch-2012:week50}
Shaman, J., Karspeck, A., and Lipsitch, M. (2012b).
\newblock Week 50 influenza forecast for the 2012-2013 {U.S.} season.
\newblock {\em ArXiv}, 1212.5750 [q-bio.PE].

\bibitem[Shaman et~al., 2012c]{Shaman-Karspeck-Lipsitch-2012:week51}
Shaman, J., Karspeck, A., and Lipsitch, M. (2012c).
\newblock Week 51 influenza forecast for the 2012-2013 {U.S.} season.
\newblock {\em ArXiv}, 1212.6678 [q-bio.PE].

\bibitem[Shaman et~al., 2013]{Shaman-Karspeck-Lipsitch-2013:week52}
Shaman, J., Karspeck, A., and Lipsitch, M. (2013).
\newblock Week 52 influenza forecast for the 2012-2013 {U.S.} season.
\newblock {\em ArXiv}, 1301.1111 [q-bio.PE].

\end{thebibliography}

%%%%%%%%%%%%%%%%%%%%%%%%%%%%%%%%%%%%%%%%%%%%%%%%%%%%%%%%%%%%%%%%%%%%%
% FIGURES
%%%%%%%%%%%%%%%%%%%%%%%%%%%%%%%%%%%%%%%%%%%%%%%%%%%%%%%%%%%%%%%%%%%%%

% \begin{figure}[tbh]
% \noindent\includegraphics[width=25pc,angle=0]{select_cities_wk52ABS_modes_probabilities.eps}\\
% \caption{Ensemble mode peak week predictions initiated December 30,
%   2012, following assimilation of Week 52 observations, for 12 cities
%   plotted as a function of probability/confidence calibrated from
%   historical city, regional and national prediction accuracy.}
% \label{fig:select_wk52fore_cal} 
% \end{figure}

% \begin{figure}[tbh]
% \noindent\includegraphics[width=18pc,angle=0]{select_cities_wk52ABS_forecastsA.eps}
% \noindent\includegraphics[width=18pc,angle=0]{select_cities_wk52ABS_forecastsB.eps}\\
% \caption{Left) Histograms of the best ensemble start date trainings
%   for forecasts made beginning the start of Week 53/1 (December 30,
%   2012) for select cities.  The distributions show the ensemble spread
%   among peak predictions.} 
% \label{fig:select_wk52fore}
% \end{figure}

%%%%%%%%%%%%%%%%%%%%%%%%%%%%%%%%%%%%%%%%%%%%%%%%%%%%%%%%%%%%%%%%%%%%%
% TABLES
%%%%%%%%%%%%%%%%%%%%%%%%%%%%%%%%%%%%%%%%%%%%%%%%%%%%%%%%%%%%%%%%%%%%%

% %
\end{document}